\documentclass[%
 aip,
 amsmath,amssymb,
 reprint,%
twocolumn,pre,aps,showpacs,superscriptaddress]{revtex4-1}

\usepackage{hyperref}
\usepackage{amsmath,amssymb}
\usepackage{amsfonts,amsthm}
\usepackage{graphics}
\usepackage{graphicx}
\usepackage{dcolumn}
\usepackage{color}
\usepackage{bm}
\usepackage{epigraph}

\usepackage{booktabs}

\usepackage[normalem]{ulem}

\begin{document}
    
    \title{The effect of the excitatory feedback in anticipated synchronization and phase bistability regimes in neuronal populations}

    \author{Julio N. Machado}
    \affiliation{Instituto de F\'{\i}sica, Universidade Federal de Alagoas, Macei\'{o}, Alagoas 57072-970 Brazil.}
    
    \author{Joana M. G. L. Silva}
    \affiliation{Instituto de F\'{\i}sica, Universidade Federal de Alagoas, Macei\'{o}, Alagoas 57072-970 Brazil.}

    \author{Katiele V. Brito}
    \affiliation{Instituto de F\'{\i}sica, Universidade Federal de Alagoas, Macei\'{o}, Alagoas 57072-970 Brazil.}
    \affiliation{Instituto de F\'{\i}sica Interdisciplinar y Sistemas Complejos IFISC (CSIC-UIB), Universitat de les Illes Balears, Campus UIB, E-07122 Palma de Mallorca, Spain}

    \author{Rodrigo Pena}
    \affiliation{Department of Biological Sciences, Florida Atlantic University, FL 33458, Jupiter, Florida, USA}
    
    \author{Fernanda S. Matias}
    \affiliation{Instituto de F\'{\i}sica, Universidade Federal de Alagoas, Macei\'{o}, Alagoas 57072-970 Brazil.}

\begin{abstract}
    Anticipated synchronization (AS), in which the receiver leads the sender and the phase lag is negative, can emerge in unidirectionally coupled dynamical systems when the receiver has faster internal dynamics than the sender. In cortical-like population models, AS and bistability between AS and delayed synchronization (DS) have been reported mainly in unidirectional motifs and have been proposed as possible explanations for phase relations observed in electrophysiological recordings. Because cortical areas are often connected bidirectionally, it is important to understand how excitatory feedback from the receiver to the sender affects the emergence and persistence of anticipatory synchronization and phase-bistable regimes.
    Here, we investigate the effect of the excitatory feedback on the phase relations between two cortical-like neuronal populations. We show that AS and phase bistability are not restricted to strictly unidirectional architectures, but remain robust in the presence of reciprocal coupling. In addition, we find that the transition from AS to DS can occur through different routes within the same motif, either via a bistable regime or via zero-lag synchronization, depending on the inhibitory coupling. More generally, the model exhibits a rich repertoire of phase relations, including: positive, negative, and zero-lag phase-lockings, as well as phase bistability, and phase-drift regimes.
    These results are consistent with the diversity of phase relations reported in electrophysiological experiments and suggest that fixed structural connectivity may support rapid reconfiguration of functional dynamics without requiring structural rewiring.
\end{abstract}

\maketitle

\begin{quotation}
    Cortical areas are extensively interconnected through reciprocal anatomical projections, yet neural communication can rapidly reorganize during perception and cognition without corresponding changes in structural connectivity. How fixed bidirectional circuits generate such flexible timing relationships remains an open question. Here, we investigate how excitatory feedback modifies the phase relations between two coupled cortical-like neuronal populations. We show that anticipated synchronization and phase bistability, previously associated mainly with unidirectional sender–receiver motifs, remain robust when reciprocal excitation is introduced. Moreover, excitatory feedback can drive the transition between anticipated and delayed synchronization through distinct dynamical routes, either through phase bistability or zero-lag synchronization, depending on the local inhibitory coupling. Our results suggest that reciprocal cortical circuits can support multiple functional leader–follower configurations within the same structural architecture, providing a possible mechanism for the rapid reorganization of inter-areal communication observed during cognitive processing.
\end{quotation}

   \section{\label{introduction} Introduction}

Behavior and perception strongly depend on flexible communication between cortical areas~\cite{palmigiano2017flexible,kohn2020principles}. This requires that an almost fixed anatomical connectivity should allow flexible changes from one functional connectivity pattern to another~\cite{Battaglia12}, which could be related to differences in phase relations~\cite{Maris13,Maris16}. Moreover, these changes should occur on timescales relevant to behavior.
In fact, diverse phase relations have been reported in many experiments related to attention and perception~\cite{Dotson14,Brovelli04,Salazar12,carlos2020anticipated}. 
Bistable phase differences in magnetoencephalography (MEG) recordings appear when participants listen to bistable speech sequences that could be perceived as two distinct word sequences repeated over time~\cite{Kosem16}.
These findings suggest that phase-bistability in cortical regions could be related to bistable perception.

Furthermore, electrophysiological findings in non-human primates have demonstrated that unidirectional causal relationships can occur with either a positive or negative phase difference between cortical regions while engaging in a cognitive task~\cite{Matias14,Montani15,Brovelli04,Salazar12}. 
Additionally, this diversity of phase relations between electrodes with unidirectional effective connection has been observed in the alpha band in human electroencephalography (EEG) recordings~\cite{carlos2020anticipated}. The counterintuitive negative phase has been explained as a regime of anticipated synchronization (AS)~\cite{Voss00,Matias14}. 
It has been recently shown that a model of two unidirectionally coupled populations can exhibit both AS and phase bistability (BI)~\cite {machado2020phase,brito2025role,brito2021neuronal}.
However, it remains unclear whether anticipated synchronization and phase bistability persist in the presence of bidirectional excitatory feedback, which is ubiquitous in cortical circuits.

\textcolor{black}{
The concept of AS was originally introduced by Voss~\cite{Voss00,Voss01b,Voss01a}, 
when two identical dynamical systems, unidirectionally coupled in a sender-receiver configuration with a delayed and negative self-feedback at the receiver, can be described by the following equations:
}
\begin{eqnarray}
\label{eq:voss}
\dot{\bf {S}} & = & {\bf f}({\bf S}(t)), \\
\dot{\bf {R}} & = & {\bf f}({\bf R}(t)) + {\bf K}[{\bf S}(t)-{\bf R}(t-t_d)], \nonumber 
\end{eqnarray}
with arbitrary continuous ${\bf f}$ and coupling matrix ${\bf K}$. 
The stable solution ${\bf R}(t)={\bf S}(t+t_d)$ describes the scenario where the receiver precedes the sender, predicting its activity with a time interval $t_d$. 
During the last decades, AS has been extensively investigated  both theoretically~\cite{Voss00,Voss01b,Voss01a,Ciszak03,Masoller01,HernandezGarcia02,Sausedo14} and experimentally~\cite{Sivaprakasam01,Ciszak09,Tang03} in physical systems.
Afterwards, it has been shown that AS can also arise due to parameter mismatches in the receiver~\cite{Kostur05,Pyragiene13,Simonov14}, and a faster internal dynamics of the receiver~\cite{Hayashi16,Dima18,Pinto19,DallaPorta19}.

Regarding brain-inspired circuits, Ciszak et al.~\cite{Ciszak03} first demonstrated AS in coupled neuron models using two FitzHugh–Nagumo neurons.
Thus, it has been shown that motifs of Hodgkin-Huxley and Izhikevich neurons coupled by chemical synapses, in which a dynamical inhibitory loop mediates the feedback, can also exhibit AS, as well as smooth transitions from AS to DS~\cite{Matias11,Pinto19}. Two neuronal populations may display diversity in phase relations  and transitions between AS and DS, influenced by synaptic dynamics~\cite{Matias14,DallaPorta19,machado2020phase} and neuronal variability~\cite{brito2021neuronal,brito2025role}. Moreover, AS has been identified in electrophysiological recordings from primates~\cite{Matias14} and in human EEG during cognitive tasks~\cite{carlos2020anticipated}, where directional interactions were evaluated using Granger causality~\cite{Matias14,Brovelli04,Salazar12,carlos2020anticipated}.

The effect of excitatory feedback from the receiver to the sender in the anticipated synchronization regime has been studied only in a simple microcircuit of three neurons~\cite{Matias16}. In this straightforward neuronal circuit, the excitatory neuron participating in the inhibitory loop is the leader, and AS is robust against the excitatory feedback. 
Moreover, this simple microcircuit does not exhibit bistability.

Here, we study the effect of excitatory feedback in anticipated synchronization and phase-bistability in a model of two neuronal populations of spiking neurons. 
Our model can exhibit phase-locking regimes with positive, negative, and zero-lag phase differences as well as phase bistability and phase drift as reported in many electrophysiological data from cortical areas~\cite{Brovelli04,Salazar12,carlos2020anticipated,Kosem16}. In Sec.~\ref{model} we describe our motif, as well as neuronal and synaptic models. In Sec.~\ref{results}, we report our results, showing that both AS and the bistable regime are robust in the asymmetric coupled motif. We also show that the excitatory feedback can induce AS-DS transition via a bistable regime or zero-lag synchronization, depending on the inhibition. Concluding remarks and a brief discussion of the significance of our findings for neuroscience are presented in Sec.~\ref{conclusions}.

\textbf{}
    
    \section{\label{model} The network motif}
         
\begin{figure}
    \centerline{\includegraphics[width=0.95\columnwidth,clip]{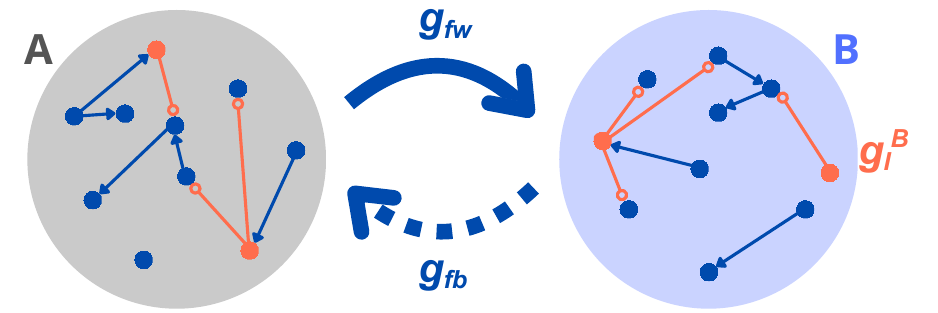}}
    \caption{ \label{fig:motif} 
        The network consists of two bidirectionally coupled cortical-like populations (A and B), each containing 500 neurons (80\% excitatory, blue; 20\% inhibitory, orange). Within each population, every neuron receives 50 recurrent synaptic inputs. Inter-population communication is mediated by 20 excitatory feedforward synapses from population A to B (conductance $g_\text{\textit{fw}}$) and 20 excitatory feedback synapses from population B to A (conductance $g_\text{\textit{fb}}$).
    }
\end{figure}

\par Although the structural (i.e., anatomical) connectivity between cortical areas is often bidirectional~\cite{Honey07}, brain functions typically require the control of inter-areal interactions on time scales faster than synaptic changes. In particular, functional and effective connectivity~\cite{Friston94} must be reconfigurable even when the underlying structural connectivity is fixed. First, different tasks require the activation of different pathways. In addition, we live in a changing environment. However, a complete understanding of how inter-areal phase coherence can be flexibly regulated at the circuit level remains lacking~\cite{Battaglia12}. To ensure that the AS and bistable regimes are not specific to cortical ensembles with unidirectional connections, their robustness is demonstrated in the presence of excitatory synaptic feedback from the receiver to the sender population.

\par Our neuronal motif is composed of two cortical-like neuronal populations: A and B (see Fig.~\ref{fig:motif}). Each one is composed of 500 Izhikevich neurons, $80\%$ excitatory and $20\%$ inhibitory~\cite{Matias14} described by:

\begin{eqnarray}
    \frac{dv}{dt} &=& 0.04v^2+5v+140-u +\sum_x I_{x}, \label{dv/dt}\\
    \frac{du}{dt} &=& a(bv-u), \label{du/dt}
\end{eqnarray}

\noindent the variable $v$ represents the neuronal membrane potential and $u$ the recovery variable, which models the activity of Na$^+$ and K$^+$ ionic currents, following the Izhikevich model~\cite{Izhikevich03}. When the neuron fires a spike ($v\geq30$~mV), $v$ is reset to $c$ and $u$ to $u+d$. Moreover, $I_x$ (see Eq.~\ref{Ix}) are the synaptic currents provided by the interaction with other neurons and external inputs. For analysis of the populations, we consider the mean membrane potential over all neurons in each population as $V^\text{A}$ and $V^\text{B}$.

\begin{table}[h!]
    \centering
    \small 
    \setlength{\tabcolsep}{10pt} 
    \renewcommand{\arraystretch}{1.2} 
    \resizebox{\linewidth}{!}{
        \begin{tabular}{c|c c c c}        
                & $a$ & $b$ & $c$ & $d$ \\ \hline \hline \\[-0.35cm]
            Exc & $0.02$ & $0.2$ & $-65+15\sigma^2$ & $8-6\sigma^2$ \\ 
            Inh & $0.02+0.08\sigma$ & $0.25-0.05\sigma$ & $-65$ & $2$ \\ 
        \end{tabular}
    }
    \caption{Izhikevich model dimensionless parameter distribution for excitatory and inhibitory neurons, where $\sigma$ is a uniformly distributed random variable on [0,1].}
    \label{tableabcd}
\end{table}

\par The dimensionless parameters in the model are randomly sampled according to \autoref{tableabcd}, where $\sigma$ is a random variable uniformly distributed on the interval $[0,1]$~\cite{Izhikevich03,Izhikevich04a}. These values account for the natural variability of neuronal dynamics in cortical networks. Excitatory neurons can be regular spiking, intrinsically bursting, or chattering, while inhibitory neurons can be fast spiking or low-threshold spiking~\cite{Izhikevich04a}. Previous studies have analyzed the effects of the variability of these parameters on synchronization regimes of unidirectionally coupled populations~\cite {brito2021neuronal,brito2025role}.

\par In the network, the connections within and between populations are given by chemical synapses. As shown in Fig.~\ref{fig:motif}, these synaptic currents are mediated unidirectionally by fast AMPA synapses for excitatory neurons and GABA$_A$ synapses for inhibitory neurons, as follows:

\begin{equation}
\label{Ix}
    I_{x} = g_{x}r_{x}(v-V_{x}),
\end{equation}

\noindent the index $x$ distinguishes excitatory ($E$) from inhibitory ($I$) synapses. We set the reversal potentials to $V_E=0$ mV and $V_I=-65$ mV. The parameter $g_x$ specifies the maximum conductance of the corresponding synapse. Moreover, the variable $r_{x}$ denotes the fraction of bound receptors, and its time evolution follows:

\begin{equation}
\label{drdt}
  \tau_x\frac{dr_{x}}{dt}=-r_{x} + D \sum_k \delta(t-t_k),\\
\end{equation}

\noindent where the summation over $k$ stands for pre-synaptic spikes at times $t_k$. We fixed $D$ at 0.05, while the decay constants were set to $\tau_{E}=5.26$ ms for excitatory synapses and $\tau_{I}=5.6$ ms for inhibitory ones~\cite{Gollo11,Matias14}.
Numerical integration was performed using the Euler method with a fixed step of $\Delta t=0.05$~ms.

\par Furthermore, to mimic an external drive, every neuron is stimulated by $n$ excitatory synapses, each modeled as an independent Poisson spike train with rate $R$. In our simulations, we used $n=1$ and $R=2400$~Hz. Unless otherwise specified, the noisy input conductances were fixed at $g_P^A = g_P^B = 0.5$~nS, for populations A and B, respectively, and a spiking frequency $R/n$, which determines the main frequency of the oscillatory activity of each population.

\par The local connectivity within each population was set to 10\% at random. Consequently, each neuron received 50 recurrent synapses from neurons of the same group. Excitatory synapses had conductance $g_E^A=g_E^B=0.5$~nS, while inhibitory synapses in population A were fixed at $g_I^A=4.0$ nS. In contrast, $g_{I}^{B}$ in population B is varied throughout the study (see Fig.~\ref{fig:motif}). Each neuron in one population receives 20 fast synapses from randomly selected excitatory neurons of the other population, with a time delay of a time step. Unless otherwise stated, the synaptic conductance of the inter-area coupling from A to B is $g_\text{\textit{fw}} = 0.5$~nS, whereas $g_\text{\textit{fb}}$ is the conductance of synapses from B to A, which are crucial to studying the role of the excitatory feedback throughout our study.

    \section{\label{results} Results}
        
\subsection{The effect of the excitatory feedback on the phase-locking regimes}
\label{sec:DS}

Starting from the situation of no excitatory feedback ($g_\textit{fb}=0$~nS), two similar populations unidirectionally coupled can synchronize in a phase-locking regime with a positive, negative, or zero phase difference~\cite{Matias14,Matias15}. The three different situations are: the delayed synchronization (DS), the anticipated synchronization (AS), and the zero-lag synchronization (ZL).  As we slowly increase the excitatory feedback $g_\textit{fb}$, we could still define populations A and B as sender and receiver as long as $g_\textit{fb}<<g_\textit{fw}=0.5$~nS.

\begin{figure}
    \centerline{\includegraphics[width=0.98\columnwidth,clip]{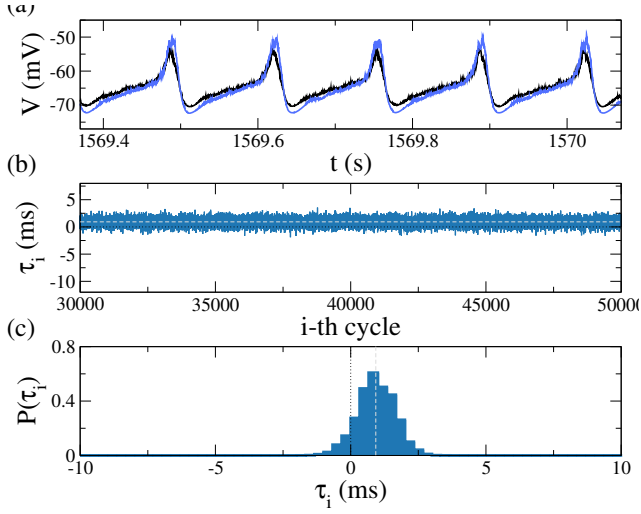}}
    \caption{
        Characterizing the usual delayed synchronization (DS) regime. (a) The mean membrane potential of populations A and B as a function of time, illustrating that a peak in the electrical activity of $V^A$ (black) typically occurs a few milliseconds before a peak of $V^B$ (royal blue); (b) The time delay in each cycle $\tau_i$ is positive in the majority of the cycles. Its average value is $\tau \approx 1$~ms; (c) The probability distribution of the time delay is unimodal. The synaptic conductances are $g_I^B = 2.0~\text{nS}$, $g_{\textit{fb}} = 0.28~\text{nS}$, and $g_{P} = 0.5~\text{nS}$.
    }\label{fig:DS} 
\end{figure}

In Fig.~\ref{fig:DS} we show an illustrative example of DS in the presence of excitatory feedback ($g_\textit{fb} = 0.28$~nS and $g_I^B = 2.0$~nS).
Fig.~\ref{fig:DS}(a) shows the mean membrane potential of each population, which is calculated as the average of all $N$ neuronal membrane potentials in that population: $V_x=\sum_{i} v_{x}^{i}/N$ ($x=$A, B).
We assume, as a crude approximation, that $V_x$ is related to the local field potential (LFP) measured in electrophysiological experiments.
We average within a sliding window of width $5-8$~ms to obtain a smoothened signal, from which we can extract 
the peak times ($t^{x}_{i}$) (where $i$ indexes the peak). The period of a given population in each cycle is thus $T^{x}_i \equiv t^{x}_{i+1}-t^{x}_{i}$. 
For a sufficiently long time series, we compute the mean period
$T_x$ and its variance. In the DS example shown in Fig.~\ref{fig:DS}, both populations oscillate with $f\simeq8$~Hz (equivalent to $T^{S}_i\simeq 125$~ms). In all those calculations, we discard the transient time.  

An important parameter for evaluating phase relations in our motif is the time delay between A and B populations. We define $\tau_i$ in each cycle as $\tau_i=t^{B}_{i}-t^{A}_{i}$,  (see the time delay $\tau_i$ in the period $i-th$ as a function of the index $i$ in Fig.~\ref{fig:DS}(b)). 
Due to this definition, if B exhibits a peak after A $\tau_i>0$, whereas $\tau_i<0$ when A exhibits a peak after B. 
Then, if $\tau_i$ obeys a unimodal distribution (Fig.~\ref{fig:DS}(c)),  we calculate $\tau$ as the mean
value of $\tau_i$ and $\sigma_{\tau}$ as its variance.  
If $T_A  \approx  T_B $
and $\tau$ are independent of the initial conditions, the populations exhibit a phase-locking regime. 

\begin{figure}
    \centerline{\includegraphics[width=0.98\columnwidth,clip]{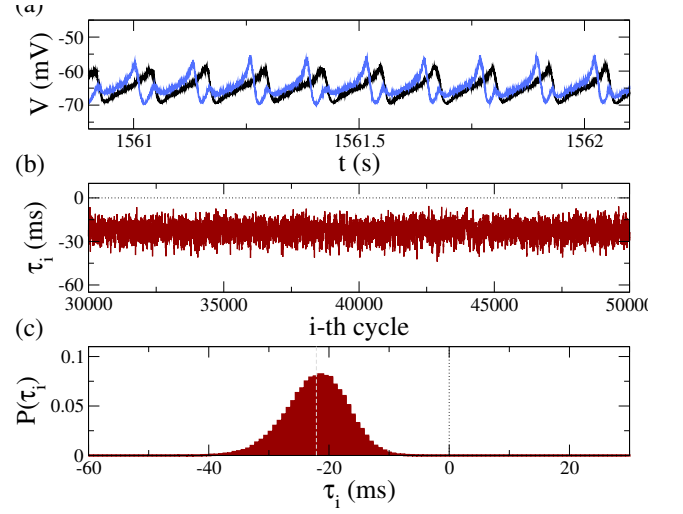}}
    \caption{
        Characterizing the anticipated synchronization (AS) regime. (a) The mean membrane potential activity of both populations, where a peak in the activity of $V^A$ (black) occurs after a peak in $V^B$ (royal blue). (b) The time delay in each cycle $\tau_i$, is negative in the majority of the cycles. (c) the probability distribution of the time delay $\tau_i$. The synaptic conductances are $g_I^B = 2.0~\text{nS}$, $g_{\textit{fb}} = 0.03~\text{nS}$, and $g_{P} = 0.5~\text{nS}$.
    }\label{fig:AS} 
\end{figure}

\begin{figure}[h]
     \centerline{\includegraphics[width=0.98\columnwidth,clip]{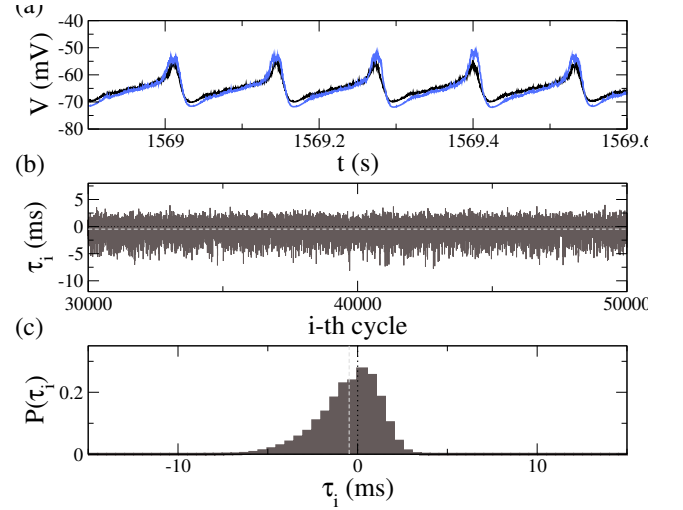}}
    \caption{
        Characterizing the zero-lag (ZL) synchronization regime ($g_I^B = 2.0$~nS, $g_\textit{fb} = 0.15$~nS and $g_{P} = 0.5~\text{nS}$) with: (a) the mean membrane potential activity of both populations $V^A$ (black) and $V^B$ (royal blue), nearly coinciding in time; (b) the time delay in each cycle $\tau_i$, is nearly Gaussian and exhibits very small variations around zero, with an average value of $\tau = -0.47$~ms \textcolor{black}{and a peak in $\tau_i = 0.15$~ms}; (c) the probability distribution of the time delay, approximately Gaussian and centered near $\tau_i = 0$~ms.
    }\label{fig:ZL} 
\end{figure}

The usual delayed synchronization regime is characterized by $\tau>0.5$~ms (see Fig.~\ref{fig:DS}).
On the other hand, the AS regime occurs for $\tau<-0.5$~ms (see Fig.~\ref{fig:AS}). 
With no loss of generality, we have arbitrarily chosen $0.5$~ms to define the intermediate region of $\tau$ close to zero, which we call zero-lag synchronization ($-0.5<\tau<0.5$~ms, see Fig.~\ref{fig:ZL}). Therefore, phase-locking regimes with the unimodal distribution of $\tau_i$ can be characterized as AS, DS, and ZL. 

In Fig.~\ref{fig:AS} we show an illustrative example of AS in the presence of a small excitatory feedback ($g_\textit{fb} = 0.03$~nS and $g_I^B = 2.0$~nS).
For these parameters, population B typically exhibits a peak before A, which is characterized by $\tau_i<0$ (Fig.~\ref{fig:AS}(b)).
Then,  $\tau_i$ obeys a unimodal distribution (Fig.~\ref{fig:AS}(c))  with $\tau<0$ as the mean
value of $\tau_i$ and $\sigma_{\tau}$ as its variance.

For larger enough values of inhibitory conductance, the unidirectional motif ($g_\textit{fb} = 0$~nS) exhibits AS. Starting from an AS regime and $g_I^B > 1.0$~nS,
as we increase the feedback conductances $g_\textit{fb}$, the system undergoes a transition from AS to DS via zero-lag synchronization 
(we provide more details below in Fig.~\ref{fig:tau_gfb-gfw0.5}(b) for $g_{I}^{B}=2.0$~nS and Fig.~\ref{fig:heatmap-gfw0.5}) This means that the system remains in a phase-locking regime, but the phase distribution goes from negative to positive values continuously.
For $g_I^B=1.0$ the transition occurs when the excitatory feedback $g_\textit{fb}=0.1$~nS is close to $20\%$ of the feedforward conductance ($g_\textit{fw}=0.5$~nS).

The reduction in the standard deviation of $\tau_i$ arises from two complementary mechanisms: (i) once the network settles into the DS attractor, the cycle-to-cycle dispersion of phase lags is smaller because in that regime, noise is filtered more efficiently. (ii) in the AS region, the system occasionally makes brief excursions to DS, producing a handful of cycles with $\tau>0$. These outliers broaden the histogram and inflate the dispersion. As the feedback strength increases, the DS state becomes dominant, then such mixed episodes disappear and the variance contracts further.


\subsection{The bistability between positive and negative phases is robust against model parameters}
\label{Bistability}

A system that spends most of its time near two well-separated regions of phase space exhibits a stationary probability density with two distinct peaks, a hallmark of bistability. A classic example is the Kramers problem, where a particle undergoes noise-driven transitions between two wells of a double-well potential. In such systems, the residence time in each state depends on the noise level.

In our case, we find that, for relatively weak inhibition at the population B ($g_{I}^{B}<<g_{I}^{A}$), the system can display a bistable regime between DS and AS.
This means that the system operates in a regime in which both DS and AS are dynamically accessible. In this regime, the instantaneous time delay 
$\tau_i$ remains positive for several oscillatory cycles, with a well-defined mean and variance, consistent with a DS state. The system then undergoes spontaneous transitions to a different dynamical regime in which $\tau_i$ becomes negative for several cycles, corresponding to an AS state.

\begin{figure}[h]
    \centerline{\includegraphics[width=0.98\columnwidth,clip]{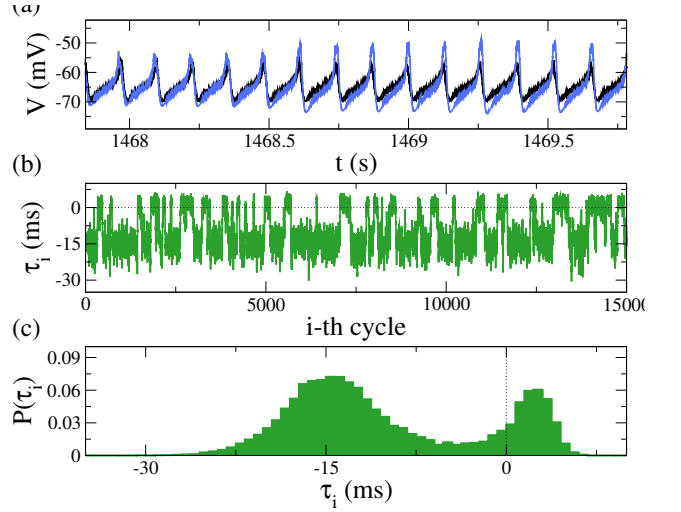}}
    \caption{
        Characterizing the bistability (BI) between the delayed and anticipated synchronization regimes. (a) the mean membrane potential activity of both populations, where $V^A$ (black) occurs after and before $V^B$ (royal blue) during different periods; (b) the time delay in each cycle $\tau_i$, which is positive for some cycles and then switches to negative values; (c) the probability distribution of the time delay follows a bimodal function with one peak at positive values of $\tau_i$ and one peak at negative values of $\tau_i$. In this example, the synaptic conductances are $g_I^B = 0.6~\text{nS}$, $g_{\textit{fb}} = 0.06~\text{nS}$, and $g_{P} = 0.5~\text{nS}$.
    }\label{fig:BI} 
\end{figure}

If the system is analyzed over short time windows, it may appear to be in either DS or AS, potentially leading to misclassification. However, over longer timescales, the system switches intermittently between these two attractors. These transitions are illustrated in Fig.~\ref{fig:BI}, which shows a typical trajectory where the system remains close to the DS attractor for several cycles before jumping to the AS regime. The reverse transition can also occur, leading to a dynamical coexistence of both states. As a consequence, the distribution of time delays is bimodal, with one peak at positive values and another at negative values (Fig.~\ref{fig:BI}(c)). Therefore, the system cannot be adequately characterized by a single average time delay.

Fig.~\ref{fig:BI}(a) shows an example of one jump from AS to DS. The system remains close to the AS-attractor for a few cycles and eventually moves to the DS region. 
The DS-events are represented by the upper states ($\tau_i>0$) in Fig.~\ref{fig:BI}(b), whereas AS-events are the lower states in the figure ($\tau_i<0$).

\begin{figure}
     \centerline{\includegraphics[width=1.00\linewidth]{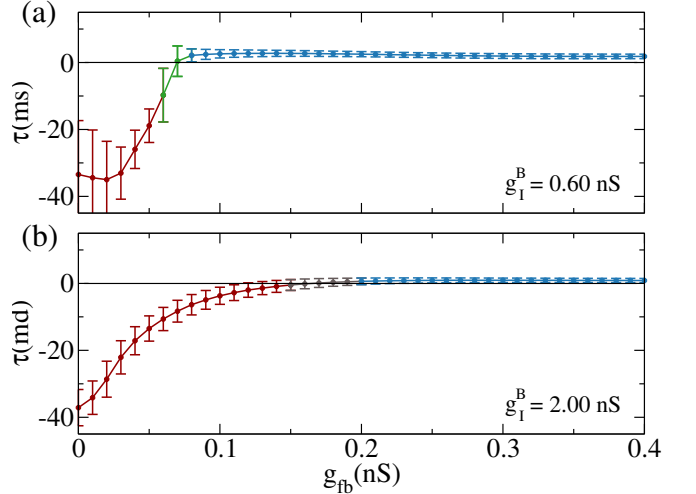}}
    \caption{
        The effect of excitatory feedback on transitions between synchronization regimes, throughout the mean time delay $\tau$. As the inhibitory coupling $g_I^B$ increases, the system loses BI behavior and shifts to transitions via ZL dynamics. The feedforward conductance used is $g_{\textit{fw}} = 0.5~\text{nS}$ and the  Poisson conductance is $g_{P} = 0.5~\text{nS}$.
    }\label{fig:tau_gfb-gfw0.5}
\end{figure}

\begin{figure}
    \centerline{\includegraphics[width=1.01\columnwidth,clip]{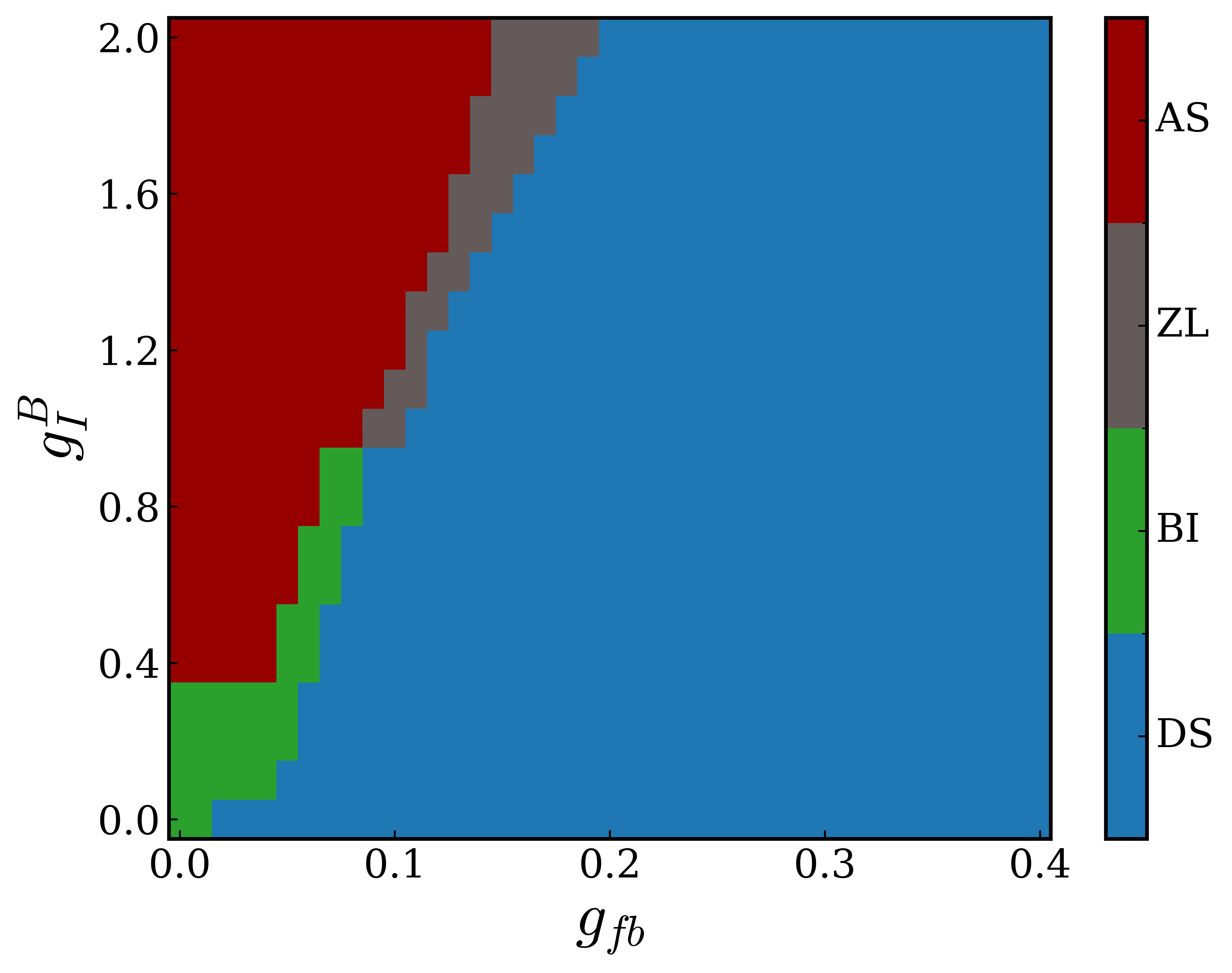}}
    \caption{ 
        For $g_\textit{fw} = 0.5$~nS and $g_{P} = 0.5~\text{nS}$, the AS–DS transition induced by excitatory feedback can occur via bistability or via zero-lag, depending on $g_I^B$, presenting BI for low values of $g_\text{fb}$ and $g_I^B$, whereas for higher values the transition occurs via ZL.
    }\label{fig:heatmap-gfw0.5}
\end{figure}

Starting from a bistable regime in unidirectional coupling ($g_{\textit{fb}}=0$), as we increase the excitatory feedback, the system undergoes a transition from bistability to a phase-locking regime with positive phase (DS). In Fig.~\ref{fig:heatmap-gfw0.5}, we show that 
for $g_{I}^{B}=0.1$~nS the bistability is lost when the excitatory feedback $g_\textit{fb}=0.05$~nS is close to $10\%$ of the feedforward conductance $g_\textit{fw}=0.5$~nS.

For intermediate inhibition values $0.4<g_{I}^{B}<0.9$~nS, the system is in an AS regime for $g_{\textit{fb}}=0$. As we increase the feedback conductance $g_\textit{fb}$, the system goes from AS to bistability and then to DS.
In Fig.~\ref{fig:tau_gfb-gfw0.5}(a) for $g_{I}^{B}=0.6$~nS we show one example of this case.
For larger values of inhibition ($g_{I}^{B}>1.0$~nS), the transition from AS to DS occurs via zero-lag synchronization (see Fig.~\ref{fig:tau_gfb-gfw0.5}(b) for $g_{I}^{B}=2.0$~nS).

In other words, by changing the feedback conductance $g_\textit{fb}$, the transition from AS to DS can be mediated by a zero-lag or bistable regime, depending on the inhibition $g_{I}^{B}$. The heatmap in Fig.~\ref{fig:heatmap-gfw0.5} illustrates the robustness of the unusual regimes AS and BI in a larger region of the parameter space  $g_{I}^{B} $ versus $g_{\textit{fb}}$.

\begin{figure}
    \centerline{\includegraphics[width=1.01\columnwidth,clip]{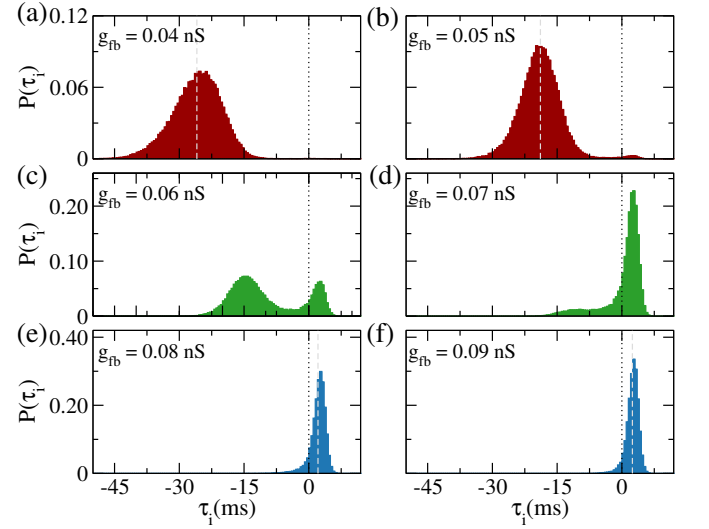}}
    \caption{Probability distributions of the time delay $\tau_i$ for the three dynamical regimes (from top to bottom): AS, BI, and DS. As $g_{\textit{fb}}$ is varied, the system exhibits a jump-like transition from AS to DS through an intermediate bistable phase. In this region, the dynamics alternates between anticipated and delayed synchronization, leading to a bimodal distribution of $\tau_i$, as illustrated in \autoref{fig:BI}. Parameters are fixed at $g_I^B = 0.6$~nS, $g_{\textit{fw}}=0.5$~nS, and $g_P=0.5$~nS.}
    \label{fig:transicao-BI-gi0.6}
\end{figure}
\begin{figure}
    \centerline{\includegraphics[width=1.01\columnwidth,clip]{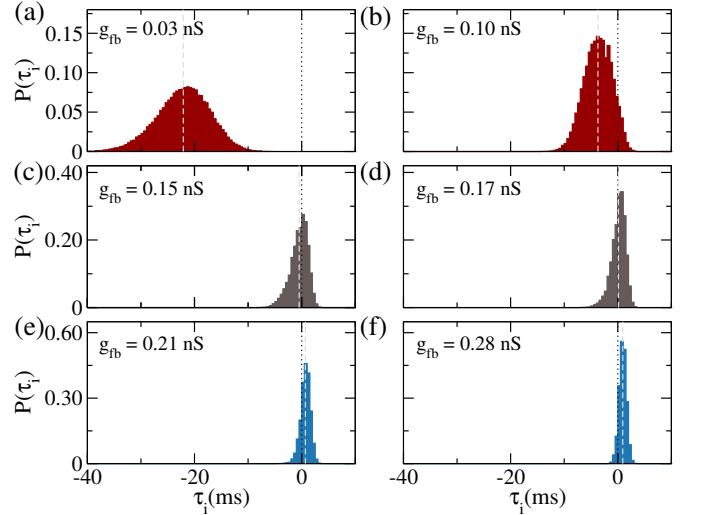}}
    \caption{Probability distributions of the time delay $\tau_i$ for the three dynamical regimes (from top to bottom): AS, ZL, and DS. As $g_{\textit{fb}}$ is varied, the system undergoes a smooth transition from AS to DS through an intermediate zero-lag synchronization state. The peak of the distribution continuously shifts from negative to positive values of $\tau_i$, crossing zero in the ZL regime, which marks the change from anticipated to delayed synchronization. Parameters are fixed at $g_I^B = 2.0$~nS, $g_{\textit{fw}}=0.5$~nS, and $g_P=0.5$~nS.}
    \label{fig:transicao-ZL-gi2.00}
\end{figure}

Figures~\ref{fig:transicao-BI-gi0.6} and \ref{fig:transicao-ZL-gi2.00} show the evolution of the probability distribution of $\tau_i$ during the AS-DS transition through the BI and ZL regimes, respectively. These transitions correspond to the horizontal traces in Fig.~\ref{fig:heatmap-gfw0.5} at $g_I^B = 0.6$~nS and $g_I^B = 2.0$~nS, whose corresponding $\tau$ evolutions are shown in Fig.~\ref{fig:tau_gfb-gfw0.5} (a) and (b), respectively. The characteristic distributions associated with the DS, AS, ZL, and the phase bistability are illustrated in Figs.~\ref{fig:DS}, \ref{fig:AS}, \ref{fig:ZL}, and \ref{fig:ZL}, respectively.

The transition through the bistable regime is characterized by the emergence of abrupt switches between AS and DS, giving rise to a second peak in the histogram and, consequently, a bimodal distribution of $\tau_i$. In contrast, the transition through the ZL regime is continuous, with the peak of the $\tau_i$ distribution shifting smoothly from negative to positive values and crossing $\tau_i=0$ within the intermediate ZL regime.

\subsection{ The effect of the external noise}  

\begin{figure}
    \centerline{\includegraphics[width=0.98\columnwidth,clip]{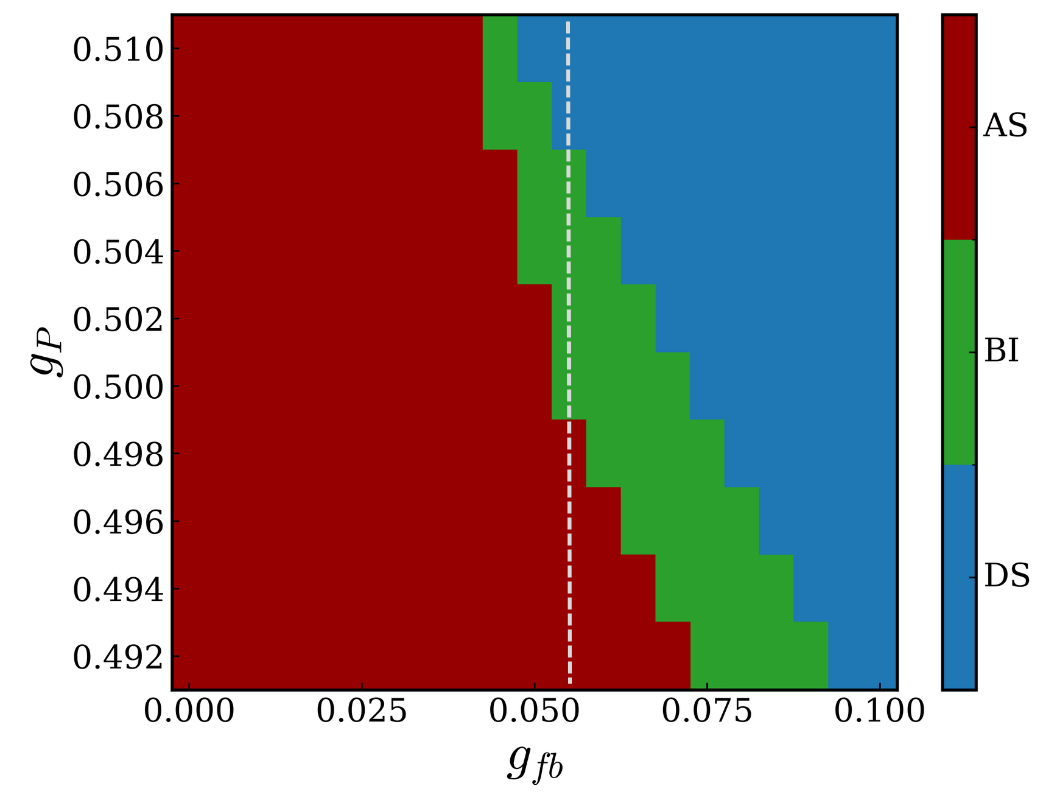}}
    \caption{
        For $g_{I}^{B}=0.6$~nS the phase diagram showing the BI-induced AS–DS transition for $g_\textit{fw}=0.5$~nS, for a small feedback conductance $g_\textit{fb}$ and with a small variation in Poisson conductance $g_P$. 
    }\label{fig:noise} 
\end{figure}

Each population receives a Poissonian input with conductance $g_P$. We investigate the effect of changes in the noise. See Fig.~\ref{fig:noise} Horizontal lines, with fixed values of $g_\textit{fb}$, show that decreasing the noise in population B promotes AS. This means that less noise facilitates the population to be the leader in phase.

\begin{figure}
\centerline{\includegraphics[width=0.98\columnwidth,clip]{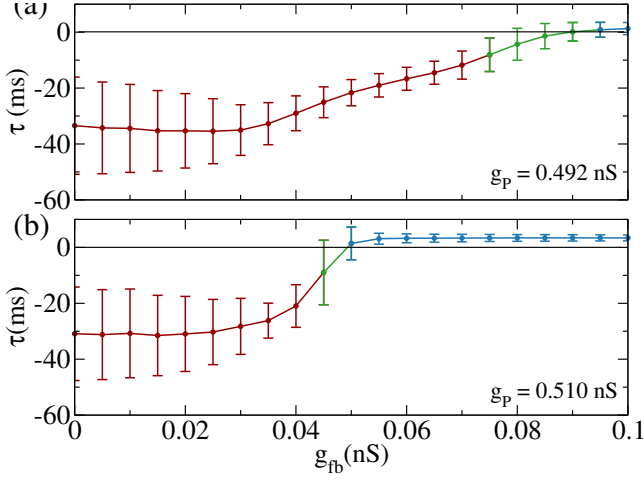}}
    \caption{The synchronization regimes as a function of the excitatory feedback conductance $g_{\textit{fb}}$, throughout the mean time delay $\tau$, for two values of the Poisson conductance $g_P$. The upper curve is characterized by an extended AS region, followed by the onset of BI dynamics and a small DS region at larger $g_{\textit{fb}}$. In contrast, the lower curve presents a broader DS regime, while preserving the same overall sequence of transitions. The feedforward conductance used is $g_{\textit{fw}} = 0.5~\text{nS}$.}
    \label{fig:tau_gfb_gP} 
\end{figure}


For a representative intermediate inhibitory coupling, $g_{I}^{B} = 0.6$~nS, we next examined how the transition between synchronization regimes depends jointly on the excitatory feedback $g_{\textit{fb}}$ and the external Poisson conductance $g_P$ at the population B (see Fig.~\ref{fig:noise}), while keeping the feedforward conductance fixed at $g_{\textit{fw}} = 0.5$~ nS. In this parameter region, the transition from AS to DS due to the excitatory feedback is still mediated by a bistable regime. Importantly, this shows that the bistable route is not restricted to a single finely tuned value of the external drive, but persists under small variations in the stochastic input.

Figure ~\ref{fig:noise} indicates that, for weak feedback, the system remains in the anticipated synchronization regime, with negative mean time delay. As $g_\textit{fb}$ increases, the network enters a parameter region in which both negative and positive lag synchronized states coexist, revealing phase bistability. For larger feedback values, the dynamics ultimately shift toward delayed synchronization, characterized by positive phase lag. Thus, in this region of parameters, the excitatory feedback does not simply symmetrize the circuit and drive it directly toward zero lag; instead, it first creates a competition between opposite leader-follower configurations before the delayed regime becomes dominant.

A similar route from AS to DS via bistability can be found for a fixed value of $g_{\textit{fb}}$ (see the dashed gray line in Fig.~\ref{fig:noise}) increasing the noise in population B. This means that lower poissonian input at population B compared to A ($g_P<0.5$~nS) facilitates AS.
Increasing the noise in population B facilitates the DS regime, whereas decreasing it favors the emergence of anticipated synchronization. This indicates that reducing the noise facilitates population B becoming the phase leader. Comparing Fig. ~\ref{fig:noise} with Fig. ~\ref{fig:heatmap-gfw0.5}, starting from a bistable regime, the effect of the Poissonian input in population B is the opposite to that of the internal inhibition. 

\subsection{The phase-drift regime}

\begin{figure}[!h]
    \centerline{\includegraphics[width=0.98\columnwidth,clip]{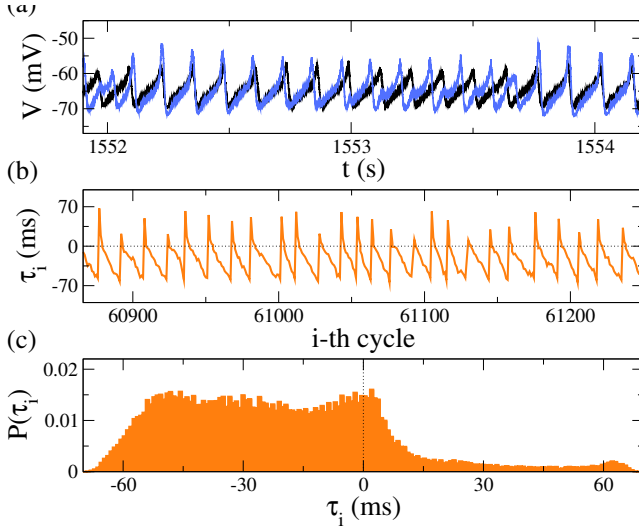}}
    \caption{
        Characterizing the phase drift ($g_\textit{fw} = 0.3$~nS, $g_I^B = 0.6$~nS, $g_\textit{fb} = 0.02$~nS and $g_{P} = 0.5~\text{nS}$) with: (a) the mean membrane potential activity of both populations, where $V^A$ (black) and $V^B$ (royal blue) do not exhibit synchronized behavior, rapidly switching between regimes; (b) the time delay in each cycle $\tau_i$, which fails to remain within a consistent regime and displays a near-periodic pattern; (c) the probability distribution of the time delay, which shows considerable variability and at times becomes almost equiprobable.
    }\label{fig:PD} 
\end{figure}

\begin{figure}
    \centerline{\includegraphics[width=1.00\linewidth]{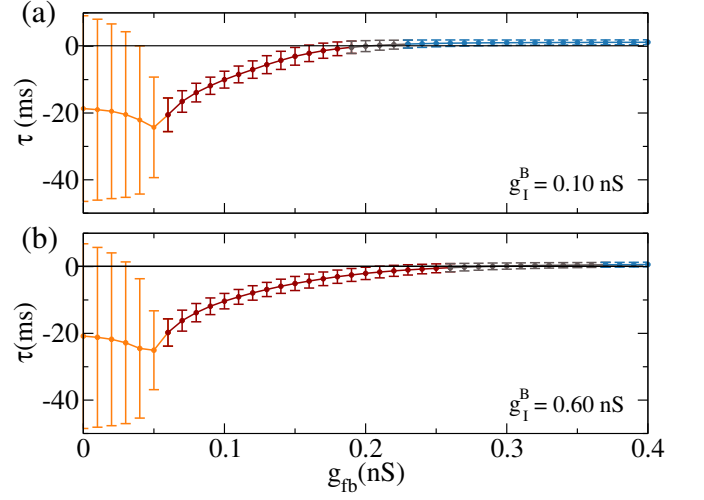}}
    \caption{
        \textcolor{black}{Effect of excitatory feedback on transitions between synchronization regimes, shown through the mean time delay $\tau$. Each curve in (a)-(d) represents a horizontal line for a fixed $g_I^B$ in Fig\ref{fig:heatmap-gfw0.3}.  Decreasing $g_{\textit{fw}}$ causes the system to oscillate in a phase-drift regime for low $g_{\textit{fb}}$, and the transition from AS to DS occurs via a zero-lag (ZL) regime. As the inhibitory coupling $g_I^B$ increases, the curves become progressively more negative, indicating stronger anticipatory behavior and the decrease of DS and ZL regimes. The conductances used here is $g_\textit{fw}=0.3$~nS and $g_{P} = 0.5~\text{nS}$.}
    }\label{fig:tau_gfb-gfw0.3}
\end{figure}

\begin{figure}
      \centerline{\includegraphics[width=1.01\columnwidth,clip]{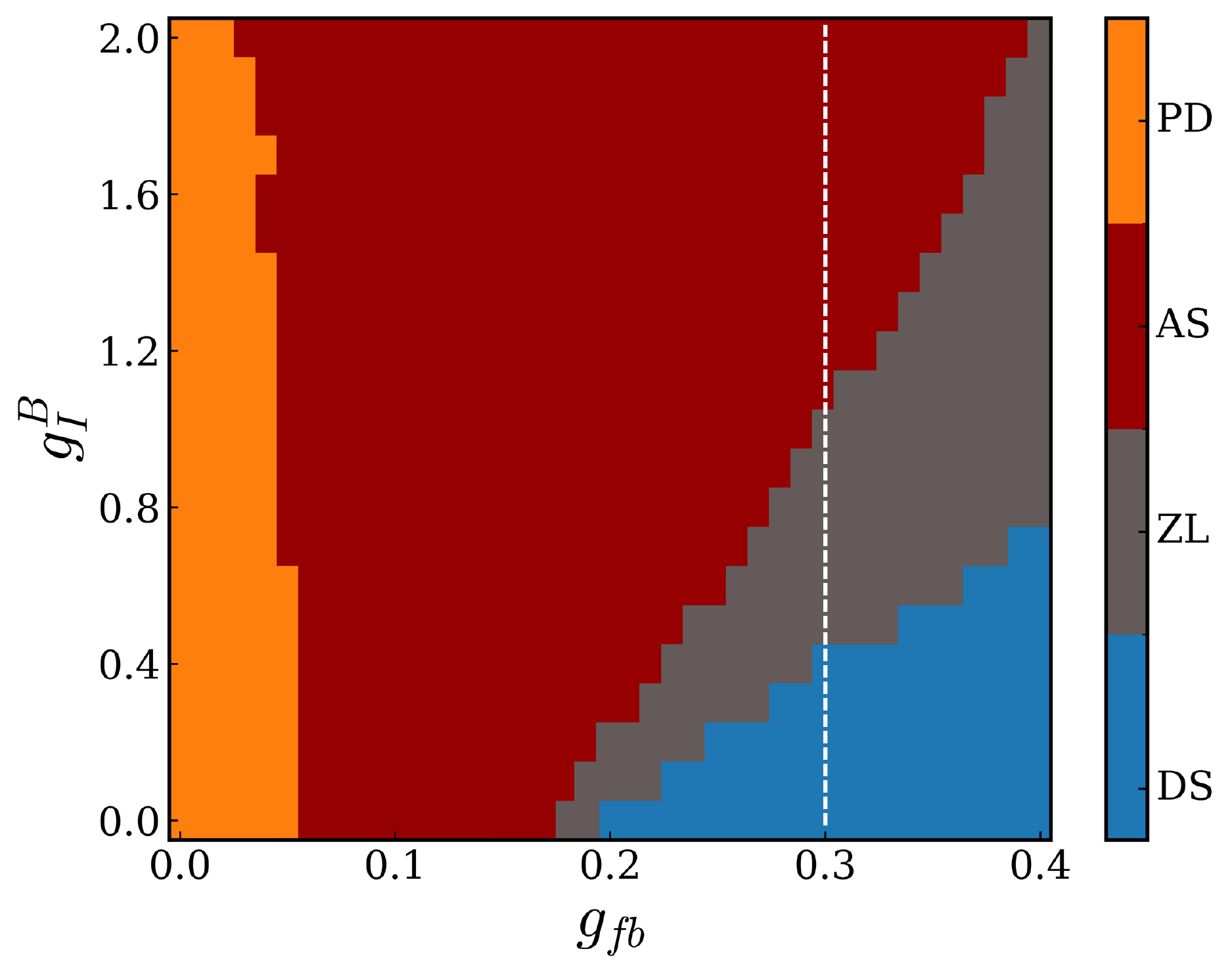}}
    \caption{
        \textcolor{black}{Two-dimensional projections of the phase diagram when varying the $g_\textit{fb}$ and $g_I^B$. For this set of parameters, the system exhibits PD when $g_\textit{fb}$ is low, beginning to synchronize in AS and transitioning to DS via ZL as $g_\textit{fb}$ increases. The conductances used here is $g_\textit{fw}=0.3$~nS and $g_{P} = 0.5~\text{nS}$.}
        }\label{fig:heatmap-gfw0.3} 
\end{figure}

A different approach is to explore the effect of the noise
when the coupling between the two populations is not large enough to ensure synchronization. When the inter-area connections ($g_\textit{fw}$ and $g_{\textit{fb}}$ are smaller than the synaptic conductances simulating the Poissonian noise $g_{P}$, the system can present a phase drift regime (PD). In contrast to the phase-locking regimes and the bistable dynamics, Fig.~\ref{fig:PD} illustrates a PD regime obtained for smaller values of the excitatory interarea connections $g_\textit{fw} = 0.3$~nS, $g_{\textit{fb}} = 0.02$~nS, $g_{I}^{B} = 0.6$~nS, and $g_P = 0.5$~nS. This parameter set corresponds to the point $g_{\textit{fb}} = 0.02$~nS on the curve shown in Fig.~\ref{fig:tau_gfb-gfw0.3}(b), which in turn is the horizontal trace at $g_I^B = 0.6$~nS in the phase diagram of Fig.~\ref{fig:heatmap-gfw0.3}. In this case, the oscillatory activity of the mean membrane potentials of the two populations does not remain locked in a stable phase relation. Instead, the relative timing between populations changes continuously over time, indicating the absence of phase-locking synchronization.

This loss of phase locking is also evident in the cycle-by-cycle delay $\tau_{i}$, which does not remain confined to a narrow interval associated with either AS or DS. Rather, $\tau_i$ evolves across a broad range of values and displays an almost periodic wandering pattern (see Fig.~\ref{fig:PD}(b)). Consistently, the probability distribution of delays becomes broad and, at times, nearly equiprobable across positive and negative values (illustrated in Fig.~\ref{fig:PD}(c)). Together, these observations identify a genuine phase-drift state, distinct from both bistability and zero-lag synchronization.

Fig.~\ref{fig:tau_gfb-gfw0.3} shows how the mean time delay $\tau$ varies with excitatory feedback $g_{\textit{fb}}$ for two values of the inhibitory conductance when the feedforward strength is reduced to $g_{\textit{fw}} = 0.3$~nS and $g_{P} = 0.5$~nS. In contrast to the stronger feedforward case shown previously, the transition from AS to DS now occurs predominantly through a zero-lag route, and for sufficiently low $g_{\textit{fb}}$ the system may also display phase drift.

For weak inhibitory coupling, the curves reveal a progression from negative delays at low feedback, through a narrow region close to zero lag, and then toward positive delays as feedback increases. As $g_{I}^{B}$ becomes larger, the curves shift downward, indicating that the anticipatory regime is strengthened and that both the delayed and zero-lag regions are progressively reduced.

The trends observed in Fig.\ref{fig:tau_gfb-gfw0.3} are summarized globally in Fig.\ref{fig:heatmap-gfw0.3}, which presents the phase diagram in the $(g_{\textit{fb}}, g_{I}^{B})$ plane for $g_{\textit{fw}} = 0.3$~nS. This heatmap makes clear that reducing the feedforward conductance reshapes the organization of dynamical regimes. In particular, the phase-drift region occupies a significant portion of parameter space at low feedback, while the zero-lag regime forms the main intermediate route separating AS from DS.

As the inhibitory conductance increases, the AS region expands and progressively invades areas that, for weaker inhibition, would support zero-lag or delayed synchronization. Conversely, the DS region becomes restricted to higher feedback values and lower inhibitory strengths. The dashed line at $g_{\textit{fb}} = 0.3$~nS in Fig.\ref{fig:heatmap-gfw0.3} represents the mutually coupled situation. In this case, increasing the internal inhibition of the population facilitates it becoming the phase leader.

    \section{\label{conclusions} Concluding remarks}
         To summarize, we investigated how excitatory feedback shapes phase relations in coupled neuronal populations. Our results show that anticipated synchronization and phase bistability, two phenomena previously emphasized mainly in unidirectional motifs, remain possible under bidirectional coupling and are qualitatively reorganized by reciprocal interactions. In particular, the emergence of AS and bistability depends on the interplay between inter-areal excitation and the local inhibitory and noisy dynamics of one of the populations. Specifically, we introduced a biologically plausible spiking population model with bidirectional connections that robustly exhibits bistability between two synchronized regimes with negative (AS) and positive (DS) phase differences. The coexistence of these regimes, together with the possibility of rapid switching between them, provides a parsimonious account of the diversity of inter-areal phase relations reported in electrophysiological recordings and of their fast modulation on cognitive timescales.

Beyond demonstrating the persistence of AS and AS-DS bistability under bidirectionality, our results also show that the transition from AS to DS can occur through distinct dynamical routes within the same circuit motif. Depending on the balance between excitation and inhibition, the system may switch through a bistable regime or pass through an intermediate zero-lag synchronization state. This flexibility suggests a possible mechanistic link to short-latency neural and perceptual responses reported in the visual system~\cite{Orban85,Nowak95,Kerzel03,Jancke04,Puccini07,Martinez14}, olfactory system~\cite{Rospars14}, and human perception~\cite{Stepp10,Stepp17}, where effective timing relationships can change rapidly without requiring modifications of the underlying anatomical wiring.

Importantly, to the best of our knowledge, this is the first population model to systematically characterize how excitatory feedback affects synchronization regimes typically associated with unidirectional coupling, including AS and AS-DS phase bistability. More broadly, these findings support a mechanistic framework in which cortical circuits with relatively fixed structural connectivity can nonetheless express multiple functional timing relationships through changes in effective coupling and local dynamics. Future work should investigate how this phase diversity scales in larger networks and how learning rules, particularly spike-timing-dependent plasticity~\cite{Abbott00,Clopath10,Matias15}, may shape, stabilize, and exploit these regimes to support adaptive cortical dynamics.
    
    \begin{acknowledgments}
        The authors thank CNPq (grants 402359/2022-4, 314092/2021-8), FAPEAL (grant APQ2022021000015), UFAL, CAPES and L’ORÉAL-UNESCO-ABC For Women In Science (Para Mulheres na Ciência) for financial support.
    \end{acknowledgments}
    %
    
    \bibliography{matias}
    
\end{document}